\documentclass[%
 reprint,
superscriptaddress,
 amsmath,amssymb,
 aps,
]{revtex4-2}

\usepackage{graphicx}
\usepackage{dcolumn}
\usepackage{bm}
\usepackage{enumerate}
\usepackage{mathtools}
\usepackage{appendix}

\usepackage{braket}
\usepackage[linesnumbered,ruled,vlined]{algorithm2e}
\DeclareMathOperator*{\argmax}{arg\,max}
\usepackage{physics}
\usepackage{subcaption}

\usepackage{xcolor}

\begin{document}

\preprint{APS/123-QED}

\title{Approaches to constrained quantum approximate optimization}

\affiliation{%
Argonne National Laboratory, 9700 S. Cass Ave., Lemont, IL 60439, USA.
}%
\author{Zain H. Saleem}%
\email{zsaleem@anl.gov}
\affiliation{%
Argonne National Laboratory, 9700 S. Cass Ave., Lemont, IL 60439, USA.
}%

\author{Teague Tomesh}
\email{ttomesh@princeton.edu}
\affiliation{%
Department of Computer Science, Princeton University, Princeton, NJ 08540, USA.
}

\author{Bilal Tariq}%
\email{bilaltar@buffalo.edu}
\affiliation{%
Argonne National Laboratory, 9700 S. Cass Ave., Lemont, IL 60439, USA.
}%

\affiliation{%
Department of Physics, University at Buffalo, SUNY, Buffalo, NY 14260, USA.
}

\author{Martin Suchara}%
\email{msuchara@anl.gov}
\affiliation{%
Argonne National Laboratory, 9700 S. Cass Ave., Lemont, IL 60439, USA.
}%


\date{\today}

\begin{abstract}

We study the costs and benefits of different quantum approaches to finding approximate solutions of constrained combinatorial optimization problems with a focus on Maximum Independent Set. In the Lagrange multiplier approach we analyze the dependence of the output on graph density and circuit depth. The Quantum Alternating Ansatz Approach is then analyzed and we examine the dependence on different choices of initial states. The Quantum Alternating Ansatz Approach, although powerful, is expensive in terms of quantum resources. A new algorithm based on a "Dynamic Quantum Variational Ansatz" (DQVA) is proposed that dynamically changes to ensure the maximum utilization of a fixed allocation of quantum resources. Our analysis and the new proposed algorithm can also be generalized to other related constrained combinatorial optimization problems.
\end{abstract}

\maketitle

\section{Introduction}
\label{sec:introduction}

Rapid progress is being made in building quantum hardware with increasing qubit counts and fidelities \cite{gambetta2017building, wright2019benchmarking, saffman2019quantum, arute2019quantum, arrazola2021quantum}. However the capabilities of quantum hardware are still limited, and have thus far been unable to demonstrate an advantage over classical processors for useful applications. Whether a useful problem can be solved on these noisy intermediate-scale quantum (NISQ) devices \cite{preskill2018quantum,dalzell2020many} remains an open question that requires developing algorithms tailored to the salient properties of the compiler and hardware stacks that support the application layer. NP-hard combinatorial optimization \cite{lucas2014ising} is an example of a class of problems that have been gaining attention due to their potential to demonstrate quantum advantage. 
Combinatorial optimization problems are also either constrained or unconstrained, depending on whether restrictions are placed on the variables, which can impact the quantum resources required to find a solution.

The Quantum Approximate Optimization Algorithm (QAOA) \cite{farhi2014quantum} is a hybrid variational algorithm that uses both classical and quantum resources to find the approximate solution to combinatorial optimization problems. QAOA is composed of an optimization loop between the quantum computer -- which executes a variational ansatz (i.e. a quantum circuit) to evaluate the objective function -- and the classical processor which updates the ansatz parameters to traverse the cost landscape.

Extensive work has been published on applying QAOA to unconstrained optimization problems such as MaxCut \cite{crooks2018performance, farhi2016quantum, wang2018quantum}; however relatively less attention has been paid to constrained problems such as Maximum Independent Set \cite{wang2020x, hadfield2018quantum, hadfield2019quantum, saleem2020max}. For these combinatorial problems, QAOA outputs solutions encoded as binary bitstrings, but in constrained optimization some bitstrings may violate the constraints and are not feasible solutions. For these problems there are two ways of imposing constraints in QAOA. (1) In the Lagrange multiplier approach, a penalty term may be added to the problem's objective function. This will convert the constrained optimization problem into an unconstrained one. QAOA will still output all possible bitstrings, but the feasible solutions will have a much higher probability of appearing in the output. A pruning step is required to find the best solution. This approach, referred to as QAOA+, was studied for the maximum independent set (MIS) problem in \cite{farhi2020quantum, farhi2020quantumA}. (2) The variational ansatz can be constructed in a way that adheres to the problem's constraints. This method, introduced in \cite{hadfield2018quantum, hadfield2019quantum} is referred to as the Quantum Alternating Operator Ansatz (QAO-Ansatz). The quantum operators that make up the QAO-Ansatz ensure that the constraints are satisfied at all times and the extremization is performed only over the space of feasible solutions. 
 
In this work we study the costs and benefits of different quantum approaches for finding the approximate solutions to constrained combinatorial optimization problems with a focus on MIS. In doing so we also propose a new algorithm that reduces the cost of the QAO-Ansatz approach by making tradeoffs between quantum and classical computational resources. We begin with the QAOA+ algorithm that can be implemented using circuits that require only one or two qubit gates since the Hamiltonian contains only nearest neighbor interactions. However its output contains both feasible and infeasible solutions and therefore requires a pruning step and lowers the probability with which the optimal solution appears in the output. The QAO-Ansatz has the advantage of not requiring any pruning and the guarantee that only feasible states with appear in the output (in the noiseless setting). These advantages come at the cost of more complex quantum circuits which are required to satisfy the problem constraints. A common example is the use of multi-control gates within the variational ansatz which require high connectivity. To tackle the increased quantum resource requirements of executing the QAO-Ansatz, this paper proposes a Dynamical Quantum Variational Ansatz (DQVA) that adapts its structure to maximally utilize a fixed allocation of available quantum resources.

We begin in Sections \ref{QAOA} and \ref{MIS1} by introducing the quantum approximate optimization algorithm and the maximum independent set problem. We then analyze and discuss QAOA+ and QAO-Ansatz in Sections \ref{QAOA+} and \ref{QAO-Ansatz}. For QAOA+ we study the dependence of the algorithm on circuit depth, Lagrange multiplier and the graph density and for QAOA-Ansatz we study the quantum-classical tradeoff and the dependence of the algorithm on the initial states. In Section \ref{DQVA} we present the DQVA algorithm for solving constrained combinatorial optimization problems. Section \ref{conclusions} concludes and suggests future directions. 
 
\section{Quantum Approximate Optimization Algorithm}\label{QAOA}
QAOA works by converting a classical optimization problem formulation into the problem of characterizing a quantum operator. The graph dependent classical objective function $C(\textbf{b})$ which we are looking to optimize is defined on n-bit strings $\textbf{b} = \{b_1,b_2,b_3 \dots b_n\} \in \{0,1\}^n$. It can be written as a quantum operator diagonal in its computational basis as,

\begin{equation}\label{qoperator}
   C_{obj} | b \rangle = C(\textbf{b}) |b \rangle.
\end{equation}

The main goal in QAOA is to prepare the ground state of this operator using a variational ansatz. The variational ansatz contains three main components:

\begin{enumerate}
    \item Initial State $|s\rangle$: This is the state on which we act with unitary operators to build our variational ansatz. 
    
    \item Phase Separator Unitary $e^{i \gamma C}$: $C$ is a diagonal operator in computational basis related to the problem at hand and is usually the same as the objective operator. This unitary plays the role of a phase separator between successive applications of the mixing operator. $\gamma$ is a variational parameter with values between $[0,2\pi]$.

    \item Mixing Unitary $e^{i \beta M}$: $M$ can be non-diagonal in the computational basis, and its job is to mix the states among each other during the optimization. $\beta$ is also a variational parameter with values between $[0,2\pi]$.
\end{enumerate}

The variational ansatz is made by combining these three components to produce the state
\begin{equation}
|\psi_p(\boldsymbol{\gamma},\boldsymbol{\beta} )\rangle = e^{-i\beta_p M}e^{-i\gamma_p C}\dots e^{-i\beta_1 M}e^{-i\gamma_1 C}|s\rangle ,   
\end{equation}
where $p$ controls the number of times the unitary operators are applied.  The expectation value of $C_{obj}$ with respect to this variational state,
\begin{equation}
E_p(\boldsymbol{\gamma},\boldsymbol{\beta})=\langle  \psi_p(\boldsymbol{\gamma},\boldsymbol{\beta} )|C_{obj}|\psi_p(\boldsymbol{\gamma},\boldsymbol{\beta} )\rangle , 
\end{equation}
is evaluated on a quantum computer. These values are passed to a classical optimizer whose job it is to find the optimal parameters that extremize $\max_{\boldsymbol{\gamma},\boldsymbol{\beta} }E_p(\boldsymbol{\gamma},\boldsymbol{\beta})$. Since the eigenstates of $C_{obj}$ are computational basis states, this maximization is achieved for the states corresponding to the solutions of the original optimization problem. 

\section{Maximum Independent Set}\label{MIS1}

\begin{figure}[h]
\minipage{0.4\textwidth}
  \includegraphics[width=7.0cm, height= 2.4cm]{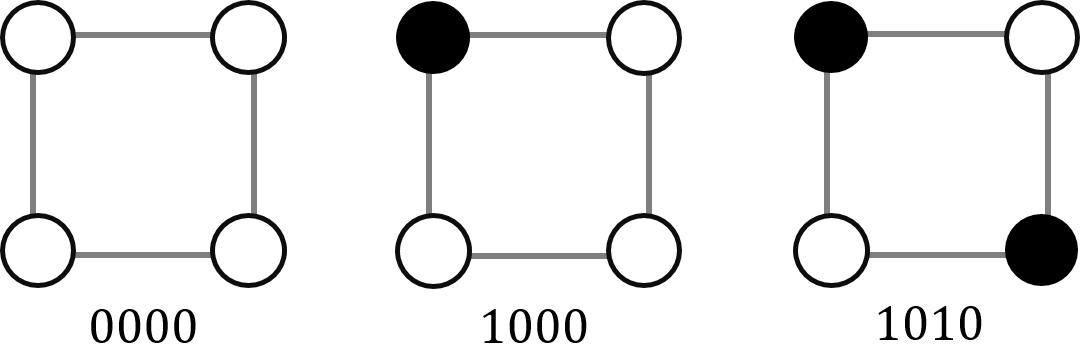}
\endminipage\hfill
   \caption{Square ring graph. The  three strings 0000, 1000, and 1010 represent the three possible independent sets of size 0, 1, and 2, respectively. 1010 (or equivalently 0101) is the string representing the maximum independent set of the graph.}\label{MIS}
\end{figure}

The Maximum Independent Set problem is one of Karp's 21 NP-hard \cite{karp1972reducibility} computational problems. This is the problem considered in this work and it is formulated as follows. Consider a graph $G = (V,E)$ where $V$ is the set of nodes of the graph and $E$ is the set of edges. A subset $V'$ of $V$ is represented by a vector $\textbf{x} = (x_i) \in \{0,1\}^{|V|}$, where $x_i = 1$ indicates that node $i$ is in the subset and $x_i = 0$ indicates that it is not included. A subset $\textbf{x}$ is called an {independent set} if no two nodes in the subset are connected by an edge: $(x_i, x_j) \neq (1,1)$ for all $(i,j) \in E$. The maximum independent set is the independent set, $\textbf{x}^*$ with the largest number of nodes. This optimization problem has the following integer programming formulation:
\begin{eqnarray}
    &&\text{maximize} \sum_{i \in V}\textbf{x}_i, \nonumber \\
    && \text{subject to} \;\; x_i x_j \neq 1 \;\; \text{where} \;\; (i,j) \in \text{E}.
\end{eqnarray}
Fig.~\ref{MIS} shows a pictorial representation of the problem. MIS is an NP-hard problem, and the best that we can do is to find an approximate solution. To find this approximation using QAOA, we start by encoding all possible states of the graph (i.e. all bitstrings $\textbf{x} \in \{0,1\}^{|V|}$) within the Hilbert space of the qubits and relating the classical objective function to a quantum operator via Eq. \ref{qoperator}. After this encoding we proceed with the variational optimization to find the optimal ansatz parameters.

\begin{figure*}[t]
    \centering
    \includegraphics[width=\textwidth]{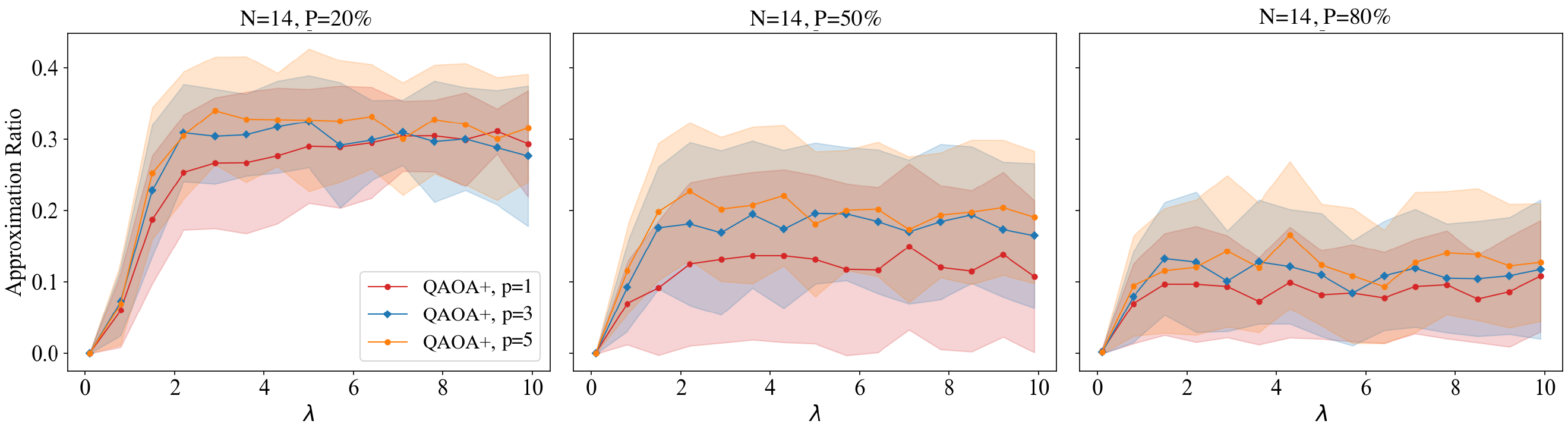}
    \caption{Dependence of QAOA+ on circuit depth $p$, edge probability $P$, and Lagrange multiplier $\lambda$. Each data point is an average over 50 Erd\"{o}s-Renyi graphs G(N,P) with $N$ the size of the graph and $P$ the edge probability. The shaded regions cover one standard deviation from the mean.}
    \label{fig:qaoa+}
\end{figure*}

\section{Penalty Term Optimization: QAOA+}\label{QAOA+}
Finding the maximum independent set of a graph $G=(V,E)$ with $N=|V|$ nodes is equivalent to finding the state with the maximum Hamming weight subject to the constraint that no two nodes (encoded as qubits) in the state share an edge. In QAOA+, this constraint is imposed at the level of the objective function by adding a penalty term that decreases the value of the objective function whenever two nodes have a common edge. The objective operator is therefore given by
\begin{equation}
    C_{obj}=H - \lambda \; C_{pen} = \sum_{i\in V} b_i - \lambda \sum_{ i,j \in E} b_i b_j ,
\end{equation}
where
\begin{equation}
    b_i= \frac{1}{2} \left(1- Z_i\right)
\end{equation}
and $Z_i$ is the Pauli-Z operator acting on the $i$-th qubit. $H$ gives us the Hamming weight of the state it acts on, and $C_{pen}$ is the term that penalizes the cost every time two neighbors are in the $\ket{1}$ state. The penalty factor $\lambda$ controls the contribution coming from the penalty term. This effectively reduces our constrained problem to an unconstrained one in the sense that now the optimization will be performed over all the bit strings $\{0,1\}^N$ during the variational optimization. 

The three components of QAOA+ are as follows:
\begin{enumerate}
    \item Initial State: $\ket{s} = \ket{+}^{\otimes N}$, where we have $\ket{+} = \frac{1}{\sqrt{2}}(\ket{0} + \ket{1})$ in the computational basis.
    \item Phase Separator Unitary: The phase separator unitary, $U_{C}(\gamma) \coloneqq e^{i \gamma C_{pen}}$, is constructed using the penalty term and parameterized by the angle $\gamma$.
    \item Mixing Unitary: The mixing unitary, $U_{M}(\beta) \coloneqq e^{i \beta \sum_i X_i}$, is parameterized by the angle $\beta$ and is composed of $X_i$ which is the Pauli-X operator acting on the $i$-th qubit.
\end{enumerate}
Together, these form the variational ansatz
\begin{equation}
    \ket{\psi_p(\boldsymbol{\gamma}, \boldsymbol{\beta})} = U_M(\beta_p)U_C(\gamma_p) \dots U_M(\beta_1)U_C(\gamma_1) \ket{s}.
    \label{eqn:plus ansatz}
\end{equation}
All components of this algorithm require at most nearest neighbor interactions and therefore can be implemented using one and two qubit gates.  

The measure we use to quantify the performance of the algorithm at circuit depth $p$ is the approximation ratio $R_p$ and it is defined as, 
\begin{equation} \label{eqn:approx-ratio}
    R_p =  \frac{E_p^{f}(\boldsymbol{\gamma}^*,\boldsymbol{\beta}^*)}{E_{max}},
\end{equation}
where $E_p^{f}$ is the expectation value over the feasible states produced by the ansatz in Eq.~\ref{eqn:plus ansatz}, $E_{max}$ is the maximum size of the MIS of the graph and $\boldsymbol{\gamma}^*, \boldsymbol{\beta}^*$ are the optimized values of the variational parameters. When calculating $E_p^{f}$ we use the following formula which requires an extra pruning step to collect only the feasible states,
\begin{equation} \label{eqn:pruning}
    E_p^{f} = \frac{\sum_i c_i H_i}{N_{s}}.
\end{equation}
The sum is performed only over the feasible states indexed by $i$, $c_i$ is the number of times the $i$-th state was observed, $H_i$ is the Hamming weight of that state and $N_s$ is the total number of shots on the quantum computer. For small graphs such as the one we use in our simulation analysis we can easily obtain $E_{max}$, however for larger graphs this is an unknown number since MIS is an NP-Hard problem.  

In Fig. \ref{fig:qaoa+} we study the QAOA+ algorithm on Erd\"{o}s-Renyi graphs with $N=14$ nodes and edge probabilities $P=20,50,80\%$, and we observe the dependence of the approximation ratio on the circuit depth, the graph density and the Lagrange multiplier $\lambda$. These experiments were performed on a noiseless simulator and even then we can see that the overall approximation ratio is quite low for all these experiments. This is because of the way that Eq.~\ref{eqn:pruning} accounts for the pruning step that QAOA+ requires. The output ($N_s$ total bitstrings) contains both feasible and infeasible solutions, but only the feasible states are included in the summation in Eq.~\ref{eqn:pruning}. These results were obtained with a modest circuit depth and only require the use of one- and two-qubit gates. The low quantum cost of running this algorithm is one of its major benefits, and,  as expected, the approximation ratio increases with the circuit depth $p$. We also find that there is a threshold value of the Lagrange multiplier $\lambda$ beyond which the approximation ratio becomes relatively constant. It is an interesting observation that the approximation ratio decreases for graphs that are more dense. It is possible that the proportion of infeasible to feasible states produced by QAOA+ increases with the graph density.

\section{Quantum Alternating Operator Ansatz}\label{QAO-Ansatz}

\begin{figure}[t]\label{graphs}
\minipage{0.4\textwidth}
  \includegraphics[width=4.5cm, height= 5.0cm]{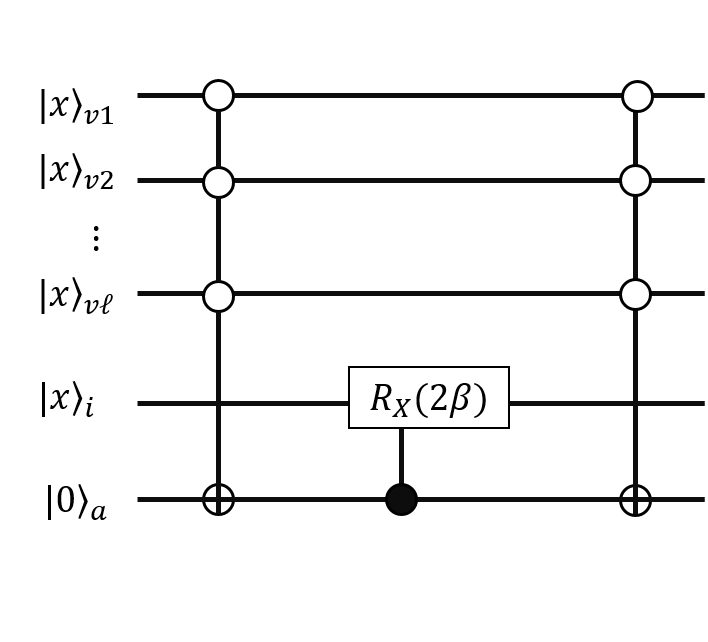}
\endminipage\hfill
   \caption{The partial mixers $V_i$ are implemented on a digital quantum computer via two multi-controlled Toffoli gates. $|x\rangle_i$ is the qubit on which we are applying the partial unitary, and $|x\rangle_{v_1}$ to $|x\rangle_{v_l}$ are the neighbors of the $i$th qubit. We can throw away the ancilla $|0\rangle_a$ after applying $V_i$. }\label{image1}
\end{figure}

\begin{figure}[t]
    \centering
    \includegraphics[width=\columnwidth]{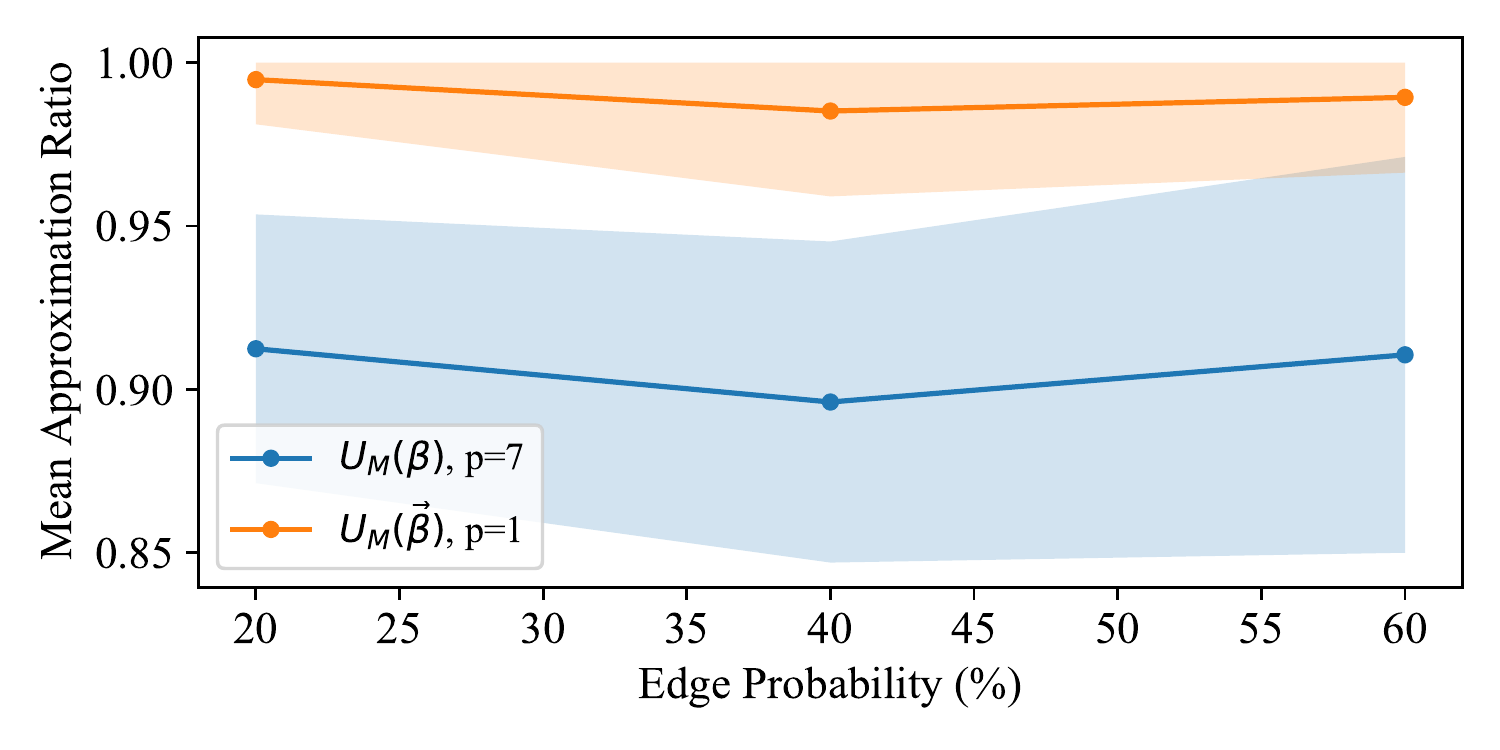}
    \caption{Comparison of the QAO-Ansatz with $U_M(\beta)$ vs $U_M(\vec{\beta})$. Averaged over 25 random 14-node Erd\"{o}s-Renyi graphs with increasing edge probability. The shaded regions denote one standard deviation from the mean.}
    \label{fig:dqva_performance}
\end{figure}

\begin{figure*}[t]
    \centering
    \includegraphics[width=\textwidth]{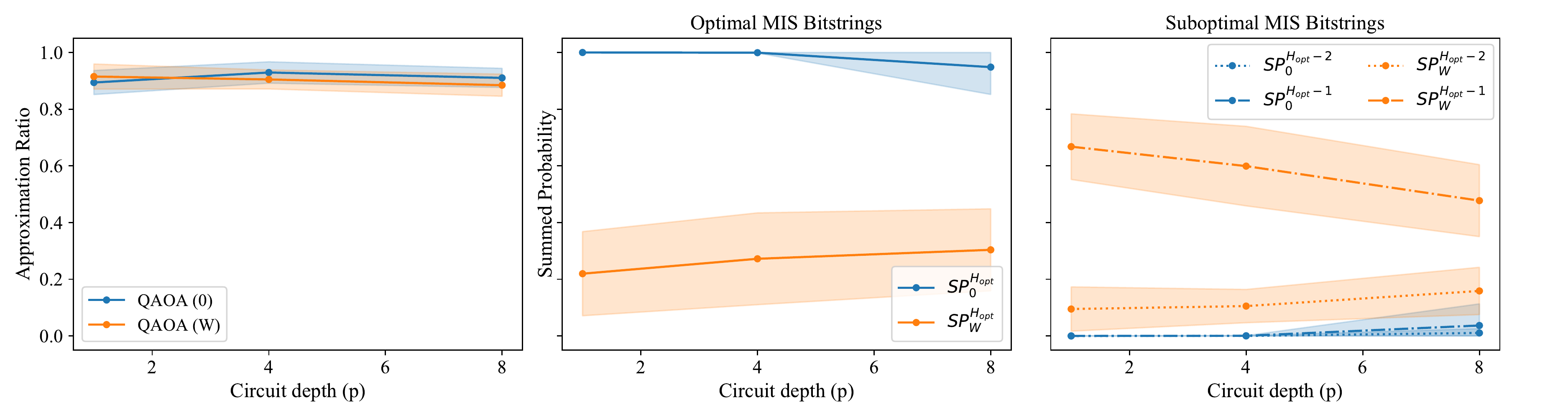}
    \caption{All data in this figure was collected by repeatedly executing each algorithm 40 times on 25 randomly generated Erd\"{o}s-Renyi graphs with 14 nodes and 20\% edge probability. All error bars denote one standard deviation from the mean. \textbf{Left}: Average approximation ratio. \textbf{Middle}: Summed probability of optimal MIS bitstrings. \textbf{Right}: Summed probability of sub-optimal MIS bitstrings}
    \label{fig:qaoa}
\end{figure*}

MIS can also be solved using the quantum alternating operator ansatz (QAO-Ansatz) where the constraints are imposed within the variational ansatz instead of the objective function. To accomplish this, the ansatz must be built in such a way that we never leave the set of feasible states during the variational optimization. The objective function here is simply the Hamming weight operator,
\begin{equation}
    C_{obj}= H = \sum_{i \in V} b_i .
    \label{qaoa objective}
\end{equation}
The three components of the QAO-Ansatz can be combined to form a variational ansatz similar to that shown in Eq.~\ref{eqn:plus ansatz}:
\begin{enumerate}
    \item Initial State: This can be any feasible state or a superposition of feasible states. An interesting choice in this regard~\cite{wang2020x} is the state,
\begin{equation}
  W= \frac{1}{\sqrt{N}}\left(|100..0\rangle + |010..0\rangle + \cdots |000..1 \rangle  \right),
\end{equation}
which is the superposition of all the single-node feasible states and can be prepared with $\mathcal{O}(N)$ CNOT gates \cite{schon2007sequential,wang2009efficient}.
    
    \item Phase Separator Unitary: The unitary $U_{C}(\gamma) \coloneqq e^{i \gamma H}$ is parameterized by the angle $\gamma$ and incorporates the objective function (Eq.~\ref{qaoa objective}).
    
    \item Mixing Unitary: Let $U_{M}(\beta) \coloneqq \prod_i e^{i \beta M_i}$, where $M_i = X_i \bar{B} $ and we have defined
    \begin{equation}
        \bar{B} \coloneqq  \prod_{j=1}^{\ell} \bar{b}_{v_j}, \;\;\;\; \bar{b}_{v_j}=\frac{1+Z_{v_j}}{2},
    \end{equation}
    where $v_j$ are the neighbors and $\ell$ is the number of neighbors for the $i$-th node. We can also write the mixer as
    \begin{eqnarray}
    U_M(\beta)=\prod_{i=1}^{N}V_i(\beta)= \prod_{i=1}^{N} \left(I + ( e^{-i\beta X_i}-I) \;  \bar{B} \right),  
    \end{eqnarray}
    where we have used $\bar{b}_{v_j}^2=\bar{b}_{v_j}$.
    The unitary mixer above is a product of $N$ partial mixers $V_i$ and in general they may not all commute with each other $[V_i , V_j] \neq 0 $. The partial mixers are executed on a digital quantum computer using the circuit shown in Fig.~\ref{image1}. We have the freedom of choosing the ordering of these partial mixers in the product and different orderings can have distinct outputs for different problem instances. Our variational ansatz therefore is defined up to a permutation 
    \begin{equation}
        U_M(\beta) \simeq \mathcal{P}\big(V_1(\beta)V_2(\beta)\cdots V_N(\beta)\big), 
    \end{equation}
    where $\mathcal{P}$ is the permutation's function of labels from 1 to $N$ and the $\simeq$ symbol represents that the mixer is defined up to permutations of $V_i$.
\subsection{Quantum-Classical Tradeoff}
We may also allow the QAO-Ansatz mixing unitary to be parameterized by a vector of angles $\vec{\beta}$ such that each partial mixer can have a different classical variable as a parameter:
    
\begin{equation}
U_M(\vec{\beta}) \simeq \mathcal{P}\big(V_1(\beta_1)V_2(\beta_2)\cdots V_N(\beta_N)\big). 
\end{equation}
\end{enumerate}

In Fig.~\ref{fig:dqva_performance} we compare the performance of the QAO-Ansatz with $U_M(\beta)$ vs $U_M(\vec{\beta})$ on 14-node connected Erd\"{o}s-Renyi graphs. We again use the approximation ratio with respect to the optimal MIS (Eq.~\ref{eqn:approx-ratio}), but we see much higher values in Fig.~\ref{fig:dqva_performance} compared to Fig.~\ref{fig:qaoa+} because every bitstring output by the QAO-Ansatz is a valid independent set.

To keep the comparison fair we ran the QAO-Ansatz with $U_M(\beta)$ using a circuit depth of $p=7$ and with $U_M(\vec{\beta})$ we restrict the circuit depth to $p=1$. This ensures that the number of classical parameters is the same between the runs with $U_M(\beta)$ and $U_M(\vec{\beta})$. Performance is significantly improved when each of the partial mixers is given an independent classical parameter. Additionally, the quantum circuits which implement the QAO-Ansatz with $U_M(\vec{\beta})$ and $p=1$ are much shallower, requiring fewer multi-control Toffoli gates, than the circuits needed to run the QAO-Ansatz with $U_M(\beta)$ and $p=7$.

\subsection{Initial State Dependence}
We analyze the impact of initial state choice on the performance of the QAO-Ansatz in Fig.~\ref{fig:qaoa} where we compare the initial states $\ket{0}$ and $\ket{W}$. The approximation ratios achieved by both initial states are high and similar in comparison. However the distributions of the output states are very different. To study the distribution of the output states we calculate the summed probabilities (SP) of the output states defined as
\begin{equation}
   SP_{I}^{H}= \sum_{i \in P_{H}} p_i
\end{equation} 
where the subscript $I= \{0, W\}$ on the left side of the equation indicates the initial state, $P_{H}$ is the set of all states with Hamming weight $H$ and $p_i$ is the probabiltiy of seeing the $i$-th state. Our simulation results show that when we choose the $|0\rangle$ initial state, the summed probabilities for the optimal Hamming weight $H=H_{opt}$ states, $SP_{0}^{H_{opt}}$ is high and the summed probabilties of sub-optimal states $SP_{0}^{H_{opt}-1}, SP_{0}^{H_{opt}-2}...$ is low. However, with the $|W\rangle$ state --- while the optimal bitstrings still have the highest overall probabilities (indicated by the high approximation ratio in the left plot of Fig.~\ref{fig:qaoa}) --- we see an increased chance of measuring sub-optimal states in the output and therefore $SP_{W}^{H_{opt}-1}, SP_{W}^{H_{opt}-2}...$ are relatively high. This suggests that the probability distribution produced by the $\ket{W}$ initialization is more evenly spread over both optimal and sub-optimal solutions. This may be a desirable feature for some applications like portfolio optimization where suboptimal solutions with higher risk may be advantageous in terms of the higher returns they offer.

\section{Dynamic Quantum Variational Ansatz}\label{DQVA}
We propose a new hybrid algorithm for constrained combinatorial optimization that allows us to dynamically alter the form of the variational ansatz while utilizing a constant amount of quantum resources.
The steps given below outline the DQVA algorithm for a given circuit depth $p$; the pseudocode is shown in Alg.~\ref{alg:dqva} and an implementation is available via Github~\cite{approaches2021github}.

\textbf{Step 1: Warm Start.} Run a classical algorithm to find a collection of bitstrings representing independent sets. A trivial choice for this collection of strings can be the bit strings representing single nodes such as $|00..1\rangle$, $|01..0\rangle$, and $|00..1\rangle$. In order to take full advantage of classical resources, however, it is advantageous to use a polynomial time classical approximate algorithm to find a set of larger Hamming weight strings representing independent sets. We will use these strings as our initial states~\cite{egger2020warm}. Let us call this set $I_{cl}$ and let
\begin{equation}
    I_{cl}= \{|\textbf{c}_1\rangle, |\textbf{c}_2\rangle \cdots |\textbf{c}_n\rangle\},
\end{equation}
where $\textbf{c}_1$, $\textbf{c}_2$, and $\textbf{c}_n$ are strings obtained by running the classical algorithm. We can prepare states $|\textbf{c}_i\rangle=|c^{1}_ic^{2}_i \cdots c^{N}_i\rangle $ with $\mathcal{O}(1)$ depth circuits, where  $c^{j}_i \in \{0,1\}$ represents the $j$th bit of the string. Additionally, each of the bitstrings in $I_{cl}$ could be used to initialize parallel instances of DQVA running on separate quantum computers. 

\textbf{Step 2: Mixer Initialization.} 
Select any of these strings, say, $|\textbf{c}_1\rangle$, and alternate between $p$ applications of the mixing unitary $U_{M}^k(\vec{\alpha}_k)$ and $p$ applications of the phase separator $U_{C}^k(\gamma)_k$ to produce the variational state,
\begin{equation}
    \ket{\boldsymbol{\alpha}, \boldsymbol{\gamma}} = U_C^p(\gamma_p) U_M^p(\vec{\alpha}_p) \dots U_C^1(\gamma_1) U_M^1(\vec{\alpha}_1) \ket{\textbf{c}_1},
    \label{eqn:dqva}
\end{equation}
where $\boldsymbol{\alpha} = \{\vec{\alpha}_1, \dots \vec{\alpha}_p\}$ and $\boldsymbol{\gamma} = \{\gamma_1, \dots \gamma_p\}$.
The mixer and phase separator unitaries in Eq.~\ref{eqn:dqva} share the same structure as those defined for the QAO-Ansatz in Sec.~\ref{QAO-Ansatz} above.
Furthermore, we make use of our observation in Fig.~\ref{fig:dqva_performance} and allow the parameters for each of the partial mixers to be independent.
\begin{equation}
    U_{M}^k(\vec{\alpha}_k)= \mathcal{P}\big(V^k_1(\alpha_k^1)V^k_2(\alpha_k^2)\cdots V^k_N(\alpha_k^N)\big), \;\; k=1,2\cdots p
\end{equation}

Whenever the $j$-th bit of the initial state is one, $c^j_1=1$, we set the corresponding parameter in every partial mixer to be $\alpha_k^j=0$. For example, if $|\textbf{c}_1\rangle = |010010\rangle$, where we have 1's at the 2nd and 5th positions, then the mixer unitaries will be,
\begin{equation}
    U^k_{M}(\vec{\alpha}_k)= \mathcal{P}\big( V^k_1(\alpha^1_k)\;I_2\; V^k_3(\alpha^3_k)\;V^k_4(\alpha^4_k)\;I_5 \;V^k_6(\alpha^6_k)\big).
\end{equation} 

\textbf{Step 3: Dynamic Ansatz Update.} Variationally optimize the parameters for the choice of mixing unitaries selected in the previous step to maximize
\begin{equation}
    \expval{C_{obj}}{\boldsymbol{\alpha}, \boldsymbol{\gamma}}.
\end{equation}
If this improves the Hamming weight and we get a new state $|\textbf{q}_1\rangle$ with a Hamming weight larger than $|\textbf{c}_1\rangle$, then we replace the initial state $|\textbf{c}_1\rangle$ with $|\textbf{q}_1\rangle$. We also update the mixing unitaries such that if $q_1^j=1$, we set the $j$-th parameters of each partial mixer to zero: $\alpha_k^j=0$. In the example above let us write the new state as $|\textbf{q}_1\rangle = |010110\rangle$. Then we have
\begin{equation}
    \tilde{U}^k_{M}(\vec{\alpha}_k)= \mathcal{P}\big(V^k_1(\alpha^1_k)\; I_2\; V^k_3(\alpha^3_k) \;I_4 \;I_5 \; V^k_6(\alpha^6_k)\big).
\end{equation} 

We then repeat steps 2 and 3 using this new state and updated unitaries $\tilde{U}$. As we set more and more parameters of the partial mixers to zero, we can add another mixing layer $U_M^{p+1}$ to ensure maximal utilization of available quantum resources. We repeat this step until we can no longer increase the Hamming weight.

\textbf{Step 4: Randomization.} If we are unable to increase the Hamming weight of the state at the above step, then we randomize over the position of the partial unitaries that have not been set equal to identity. A hyperparameter $m$ is used to set an upper limit on the number of times to perform this randomization step. In our example, this randomization operation $\mathcal{R}$ may give,
\begin{equation}
    \mathcal{R} (\tilde{U}^k_{M}(\vec{\alpha}_k)) = V^k_3(\alpha_k^3)\;I_2\; V^k_6(\alpha_k^6)\;I_4\;I_5 \; V^k_1(\alpha^1_k).
\end{equation}
After the $m$ randomizations and accompanying quantum optimizations are complete, save the final state $\textbf{q}_1^{m}$.

\textbf {Step 5: Output and Repeat.}  Repeat Steps  1--4 with the other initial states $|\textbf{c}_i\rangle$, and obtain the set
\begin{equation}
    I_{qu}= \{|\textbf{q}_1^m\rangle, |\textbf{q}_2^m\rangle \cdots |\textbf{q}_n^m\rangle\}.
\end{equation}
This set has a state $|\textbf{q}_i^m\rangle$ with the largest Hamming weight which will be returned as the final result. 

\begin{algorithm}[h]
\SetAlgoLined
\SetKwData{Left}{left}\SetKwData{This}{this}\SetKwData{Up}{up}
\SetKwFunction{Union}{Union}\SetKwFunction{FindCompress}{FindCompress}
\SetKwInOut{Input}{Input}\SetKwInOut{Output}{Output}

\Input{$G = (Q,E)$, m = \# of randomizations}
\Output{Approximate MIS of $G$}
\tcc{Repeat this procedure with multiple warm start initial states $\bm{c_j}$}
init\_state $\leftarrow \bm{c_j}$\;
$MIS_{best} \leftarrow ``00...0"$\;
\For{$r \in [m]$}{
\tcc{Mixer Initialization}
mixer\_order $\leftarrow \mathcal{P}(V_1(\alpha_1)V_2(\alpha_2) ... V_n(\alpha_n))$\;
$h_{new} \leftarrow C(\text{init\_state})$\;
$h_{old} \leftarrow -1$\;
\While{$h_{new} > h_{old}$}{
  \tcc{Inner Variational Loop}
  \While{not converged}{
    $\boldsymbol{\theta} \leftarrow \text{updated\_params()}$\;
    counts $\leftarrow$ execute($U_{dqva}(\boldsymbol{\theta})\ket{\text{init\_state}}$)\;
    $E \leftarrow$\ expectation\_value($C_{obj}$, counts)\;
  }
        $h_{old} \leftarrow C(\text{init\_state})$\;
  $h_{new} \leftarrow \max_s([C(s) \text{ for } s \text{ in counts}])$\;
  init\_state $\leftarrow \argmax_s{([C(s) \text{ for } s \text{ in counts}])}$\;
  \If{$h_{new} > C(MIS_{best})$}{
    $MIS_{best} \leftarrow$ init\_state\;
  }
}
}
 \KwRet $MIS_{best}$
 \caption{Dynamic Quantum Variational Ansatz (DQVA)}
 \label{alg:dqva}
\end{algorithm}

The DQVA algorithm takes maximum advantage of existing classical polynomial time MIS approximation algorithms with the warm start which has been shown to improve quantum optimization~\cite{egger2020warm}. Moreover, by ``dynamically" turning off and on parameters we take maximum advantage of the available quantum resources. 

\begin{figure}[t]
    \centering
    \includegraphics[width=\columnwidth]{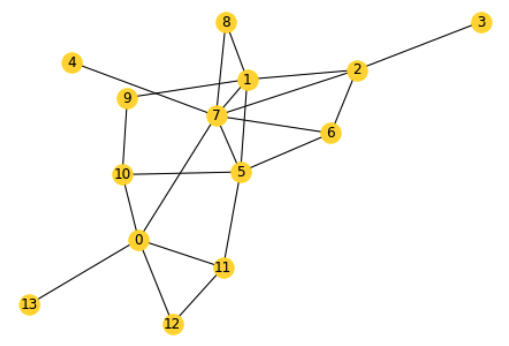}
    \caption{Randomly sampled Erd\"{o}s-Renyi ($N=14$, $P=0.2$) graph used to make Fig~\ref{fig:dqvaoverrounds}.}
    \label{fig:examplegraph}
\end{figure}

\begin{figure}[t]
    \centering
    \includegraphics[width=\columnwidth]{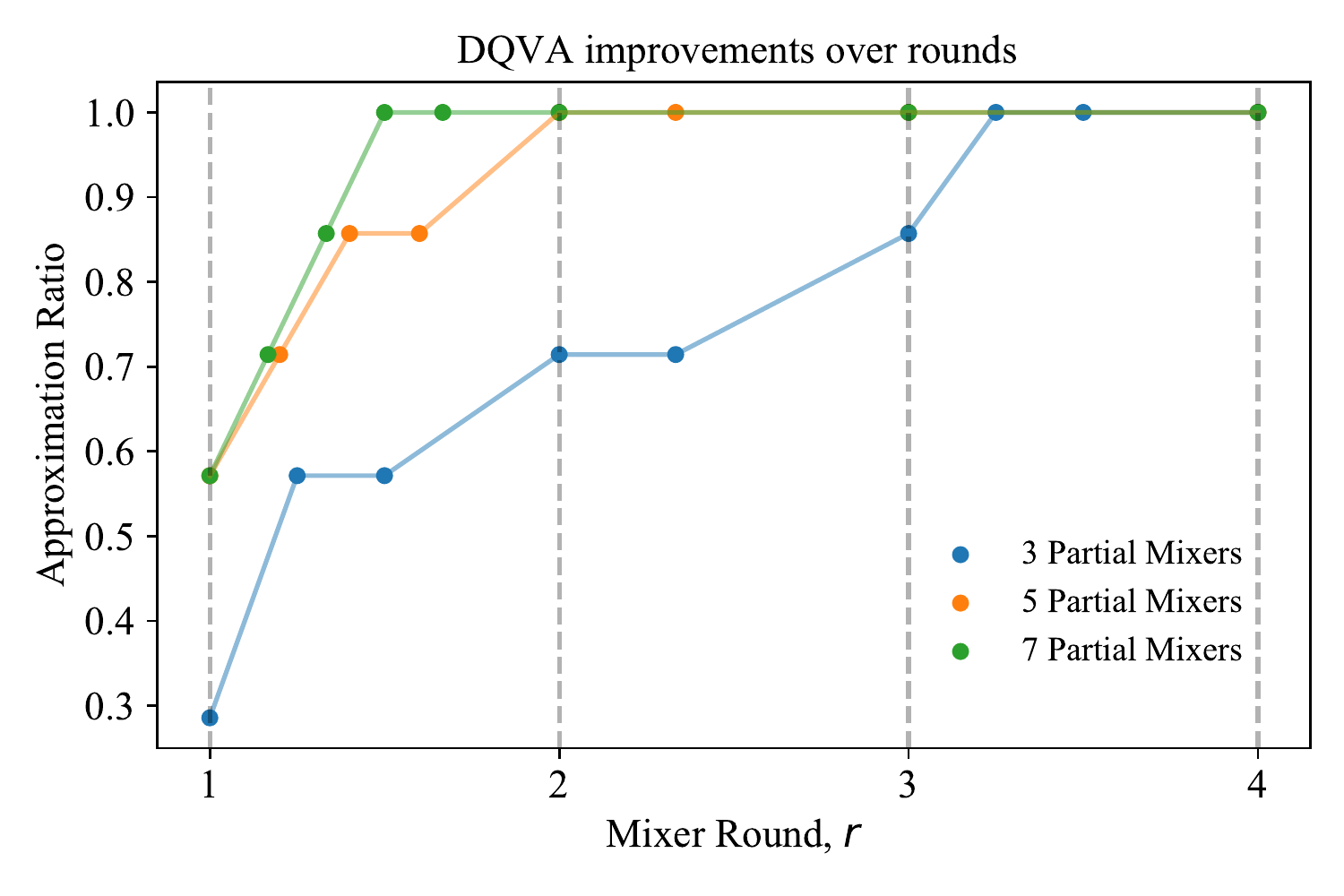}
    \caption{Improvement over dynamic ansatz update and randomization rounds for a single graph. As the size of the quantum resource budget is increased and the ansatz utilizes more partial mixers, the speed with which DQVA converges to the optimal solution increases.}
    \label{fig:dqvaoverrounds}
\end{figure}

We demonstrate the iterative execution of DQVA on a 14-node graph (shown for reference in Fig.~\ref{fig:examplegraph}) in Fig.~\ref{fig:dqvaoverrounds}. Due to the intractable scaling of quantum circuit simulation, we are limited to small graph sizes and, therefore, we do not use the warm start step of the algorithm which will be an obviously useful step to get out local minima and for enabling parallel running of the algorithm for large graphs. Instead, we use the all zero state as the initial state in our experiments.

Another hurdle we face in demonstrating the full potential of DQVA, due to the small size of the graph, is that when all partial mixers are turned on in our ansatz we will not be able to demonstrate the improvement over rounds as we update the ansatz because we will reach the best approximation ratio in the first step. We therefore restrict ourselves to using $3,5$ and $7$ partial mixers as shown in Fig.~\ref{fig:dqvaoverrounds}. The ability to restrict the amount of quantum resources is a useful feature of the DQVA algorithm which can be used to tailor its execution to the capabilities of specific quantum devices. It can be seen in Fig.~\ref{fig:dqvaoverrounds} that as we increase the number of partial mixers we reach the optimal solution faster. 
 
\begin{figure}[t]
    \centering
    \includegraphics[width=\columnwidth]{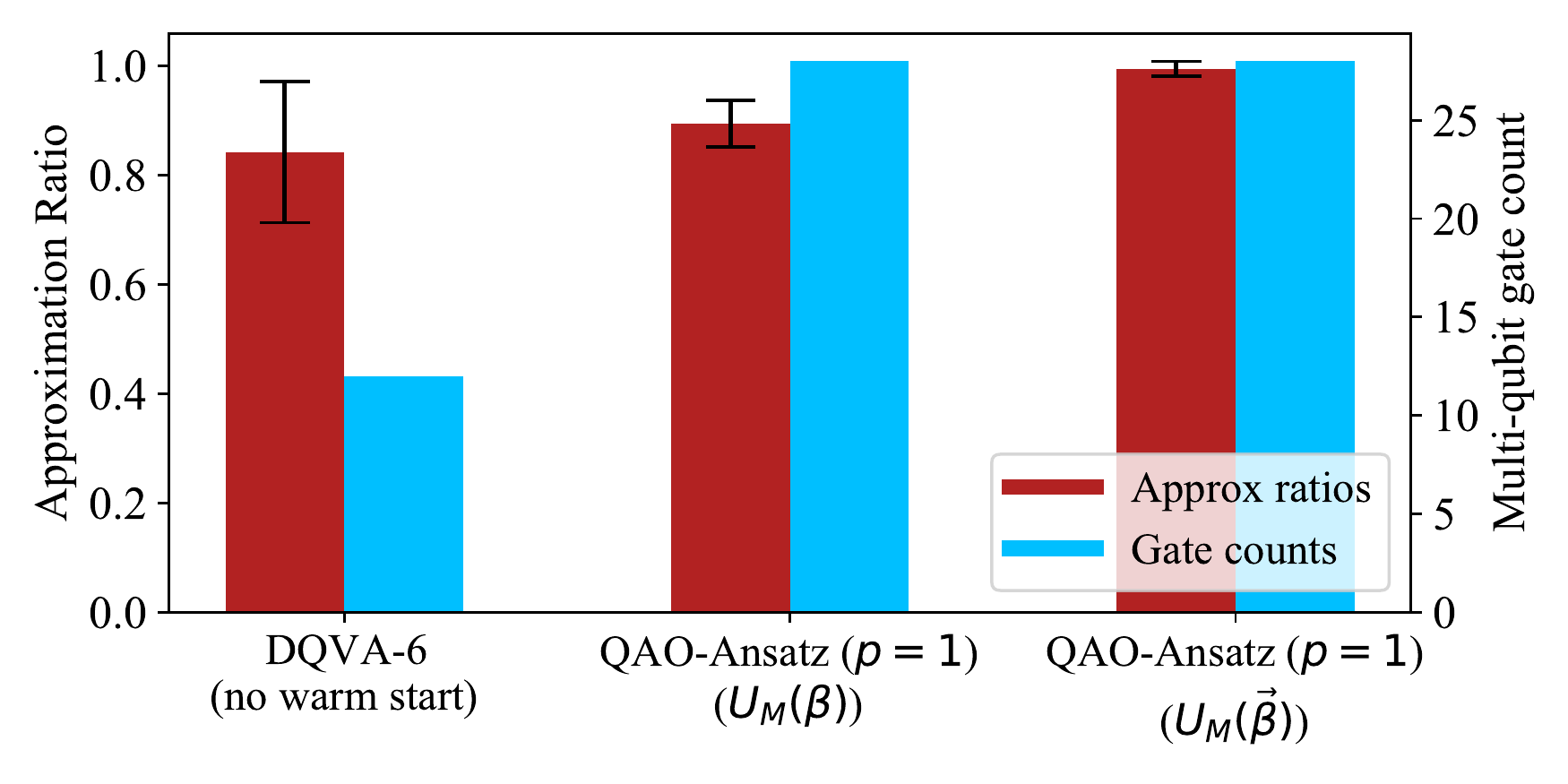}
    \caption{Performance and resource comparison between $p=1$ QAO-Ansatz ($U_M(\beta)$ and $U_M(\vec{\beta})$) and DQVA, limited to 6 partial mixers. The data is averaged over 25 random Erd\"{o}s-Renyi graphs ($N=14$, $P=0.2$), and the error bars denote one standard deviation from the mean.}
    \label{fig:resources}
\end{figure}

In Fig.~\ref{fig:resources} we highlight the resource tradeoffs made by DQVA. Starting in the $\ket{0}$ initial state, DQVA with 6 partial mixers achieves comparable performance to the QAO-Ansatz methods which use more than double the number of multi-controlled Toffoli gates. These resource savings are substantial since the decomposition of multi-qubit gates into one- and two-qubit nearest neighbor interactions can be quite expensive~\cite{vivek2009toffoli, he2017decompositions}. However, some architectures, such as neutral atom quantum computers, are especially promising because of their ability to natively implement multi-qubit gates~\cite{saffman2019quantum, saffman2016quantum}.

It is important to note that in the above execution of DQVA where we restricted the number of partial mixers, we have randomly chosen the nodes for application of the partial mixers. For a small enough graph or in the case where we apply partial mixers to all the nodes this presents no problem. However for large graphs we may end up applying partial mixers on nodes that are far apart from each other thereby reducing quantum interference between the partial mixers. In our future work we will present a strategy to apply partial mixers within a certain neighborhood as we traverse the graph. We call this strategy quantum local search. 

\section{Conclusion and Future Directions}\label{conclusions}
MIS is an important optimization problem with applications in scheduling \cite{ambuhl2009single}, inference of phylogenetic trees \cite{auyeung2004largest}, communications \cite{safar2007hard}, and portfolio optimization \cite{kalra2018portfolio}. In this work we have reviewed two existing approaches, QAOA+ and the QAO-Ansatz \cite{farhi2020quantum,farhi2020quantumA,hadfield2018quantum,hadfield2019quantum}, to constrained optimization for the MIS problem. There has been other related work on this problem as well \cite{pichler2018quantum, fernandez2020hybrid, ostroswki2020lower} but all these works are based on one of these two approaches i.e they either apply the constraint in the objective function as a penalty term or they apply it within the variational ansatz. 

In this work we have introduced the dynamic quantum variational ansatz (DQVA) that addresses the large quantum resource requirement of the quantum alternating operator ansatz. The DQVA algorithm has three main components that allow us to take advantage of the available classical and quantum resources. The first is the warm starting which takes advantage of available classical approximate polynomial time algorithms~\cite{egger2020warm}. The second component is the dynamic update of the ansatz based on the quantum outputs that the quantum approximate optimization generates. The dynamic ansatz update has similarities with reinforcement learning and could potentially benefit from its use, a problem we plan to investigate in the future.  Finally, there is the randomization that takes advantage of the permutation freedom that the mixing unitary has in choosing the ordering of the partial mixers. Using these components, we can maximize the performance of the quantum approximate optimization algorithm for constrained optimization problems using a fixed allocation of quantum resources.

For future work we plan to study the approaches presented here on classical simulators as well as on actual quantum hardware. Of interest is how much we can scale up in terms of problem size using the dynamic quantum variational ansatz. The methods we have used in our analysis and the algorithm we have suggested in this work can be applied to other constrained combinatorial optimization problems such as the Max k-Colorability, Max k-Colorable Induced Subgraph, traveling salesman problem, and max set packing. We hope to extend our study to these problems as well.

\section*{Acknowledgments}
We thank Kaiwen Gui, Brajesh Gupt, Ruslan Shaydulin, Stuart Hadfield and James Stokes for helpful discussions. This material is based upon work partly supported by the National Science Foundation under Award No. 2037984, and partly by Laboratory Directed Research and Development (LDRD) funding from Argonne National Laboratory, provided by the Director, Office of Science, of the U.S. Department of Energy under contract DE-AC02-06CH11357. T.T. is supported in part by EPiQC, an NSF Expedition in Computing, under grants CCF-1730082.

\bibliography{refs}

\end{document}